\documentclass[a4paper,11pt]{article}
\usepackage{pos}
\usepackage{hyperref}
\usepackage{graphicx}
\usepackage{amsmath,mathtools}
\usepackage{xspace}
\newcommand{\nnlojet}{NNLO\protect\scalebox{0.8}{JET}\xspace}
\def\alphas{\alpha_s}

\DeclareRobustCommand{\LO}{\text{LO}\xspace}

\DeclareRobustCommand{\N}[1]{\ensuremath{\text{N}^{#1}}}

\title{Differential N$^3$LO QCD corrections to charged current production at the LHC}

\author*[a,b]{Xuan~Chen}
\author[c]{Thomas~Gehrmann}
\author[d]{Nigel~Glover}
\author[e]{Alexander~Huss}
\author[c]{Tong-Zhi~Yang}
\author[f]{Hua~Xing~Zhu}

\affiliation[a]{Institute for Theoretical Physics, Karlsruhe Institute of Technology, 76131 Karlsruhe, Germany}
\affiliation[b]{Institute for Astroparticle Physics, Karlsruhe Institute of Technology, 76344 Eggenstein-Leopoldshafen, Germany}
\affiliation[c]{Physik-Institut, Universit\"at Z\"urich, Winterthurerstrasse 190, CH-8057 Z\"urich, Switzerland}
\affiliation[d]{Institute for Particle Physics Phenomenology, Physics Department, Durham University, Durham, DH1 3LE, UK}
\affiliation[e]{Theoretical Physics Department, CERN, 1211 Geneva 23, Switzerland}
\affiliation[f]{Zhejiang Institute of Modern Physics, Department of
  Physics, Zhejiang University, Hangzhou, 310027, China\vspace{0.5ex}}

\emailAdd{xuan.chen@kit.edu}
\emailAdd{thomas.gehrmann@uzh.ch}
\emailAdd{e.w.n.glover@durham.ac.uk}
\emailAdd{alexander.huss@cern.ch}
\emailAdd{toyang@physik.uzh.ch}
\emailAdd{zhuhx@zju.edu.cn}

\abstract{Charged current Drell-Yan production at hadron colliders is a benchmark electroweak process. A recent measurement of the W boson mass by the CDF experiment displays a large deviation from the Standard Model prediction. To enable precision phenomenology for this process, we compute the third-order (N$^3$LO) QCD corrections to the rapidity distribution in W boson production and to the transverse mass distribution of its decay products. We study kinematic regions relevant for the LHC experiments and assess the numerical magnitude of uncertainties from electroweak input parameters and parton distribution functions.}

\FullConference{%
  Loops and Legs in Quantum Field Theory - LL2022,\\
  25-30 April, 2022\\
  Ettal, Germany
}

\begin{document}

\maketitle

\section{Introduction}
Charged-current Drell-Yan production mediated by electroweak (EW) gauge bosons W$^\pm$ is a fundamental process at hadron colliders~\cite{Drell:1970wh}. 
Measurements of the inclusive and differential properties of W boson production have been performed at the Large Hadron Collider (LHC) and Fermilab Tevatron~\cite{CDF:2012gpf,D0:2012kms,ATLAS:2017rzl,LHCb:2021bjt,CDF:2018cnj,ATLAS:2019fgb,ATLAS:2019zci,CMS:2020cph}.  
Electroweak parameters of the Standard Model (SM) such as the W boson mass, decay width, weak mixing angle, and parton distribution functions (PDFs) in the proton can be determined by precision measurements of charged-current Drell-Yan production.
Recently a study by the CDF collaboration using a sample of approximately 4 million W bosons collected at the Tevatron reports the W boson mass $M_W=80433.5\pm 9.4$ MeV~\cite{CDF:2022hxs}.
The result deviates by 7 standard deviations from the SM electroweak precision fit of $M_W=80357\pm 6$ MeV~\cite{Haller:2018nnx,ParticleDataGroup:2020ssz}, which is itself in good agreement of measurements by ATLAS~\cite{ATLAS:2017rzl} and LHCb~\cite{LHCb:2021bjt} as well as the previous CDF result~\cite{CDF:2012gpf}. 
Direct measurements of the W boson mass have limited resolution due to missing energy of the neutrino in the final state charged lepton-neutrino system.
Precise measurements of $M_W$ are thus based on template fits to differential observables. 
Both the CDF~\cite{CDF:2012gpf,CDF:2022hxs} and ATLAS~\cite{ATLAS:2017rzl} analyses include distributions of final state charged lepton transverse momentum and transverse mass of the lepton-neutrino system as fitting observables. 
The measured distributions are compared with theory predictions with various input values. The $M_W$ parameter that achieves the best fit corresponds to the measured central value of W boson mass. The uncertainty associated with the measurement is a combination of statistic and systematic errors from both experimental analysis and theory predictions. 

The accuracy of theory predictions in charged current Drell-Yan production directly impact measurements of the SM electroweak parameters. 
Rapid progress has been achieved in perturbative QCD with next-to-leading order (NLO,~\cite{Altarelli:1979ub}) and next-to-next-to-leading order~(NNLO,~\cite{Hamberg:1990np}) corrections available for a while. 
Fully differential NNLO QCD corrections~\cite{Anastasiou:2003ds,Melnikov:2006kv,Catani:2009sm,Catani:2010en} are routinely used in the experimental analysis of Drell-Yan production.
The third order (N$^3$LO) corrections in QCD were accomplished recently for inclusive cross sections~\cite{Duhr:2020sdp} and for differential predictions~\cite{Chen:2022lwc}.
QCD resummation of large logarithmic corrections has been combined with fixed order predictions~\cite{Balazs:1997xd,Bozzi:2010xn,Becher:2011xn} with up to next-to-next-to-next-to-leading logarithmic (N$^3$LL) accuracy~\cite{Bizon:2019zgf,Becher:2020ugp,Isaacson:2022rts}.
These QCD corrections for the charged current Drell-Yan process combine with NLO EW~\cite{Dittmaier:2001ay,Baur:2004ig,CarloniCalame:2007cd} and mixed QCD-EW corrections~\cite{Dittmaier:2015rxo,Dittmaier:2020vra,Behring:2020cqi,Buonocore:2021rxx,Behring:2021adr} to achieve theory predictions at high quantitative accruacy. 
Impact from the parton distribution functions (PDFs) on $M_W$ determinations has been studied with NLO QCD plus parton shower corrections~\cite{Bozzi:2015hha,Farry:2019rfg,Bagnaschi:2019mzi,Gao:2022wxk} or scaled to NNLO precision~\cite{Nadolsky:2004vt,Hussein:2019kqx}.

In this study, we focus on the rapidity distribution of W boson production and the transverse mass distribution of its decay products,
which we newly compute to N$^3$LO in QCD. We identify characteristic changes at NNLO QCD accuracy in differential distributions caused by shifting EW input parameters such as $M_W$ and $\Gamma_W$, implementing an alternative Breit-Wigner parametrisation between fixed and running decay width, and switching between choices of modern PDFs. We further compare the uncertainties from above analysis at NNLO to N$^3$LO predictions in~\cite{Chen:2022lwc}. This work provides guidance for the estimation of theoretical uncertainties in future precision measurements of SM parameters associated with charged-current Drell-Yan production.

\section{Implementation}
Our prediction of the perturbative QCD corrections of charged-current Drell-Yan production up to NNLO is provided by the parton level event generator \nnlojet, which implements the antenna subtraction method~\cite{Gehrmann-DeRidder:2005btv,Daleo:2006xa,Currie:2013vh}. 
The state-of-the-art differential N$^3$LO correction is achieved through an established framework~\cite{Cieri:2018oms,Billis:2019vxg,Chen:2021vtu,Chen:2022lwc} using the $q_T$-subtraction formalism~\cite{Catani:2007vq}.
It requires the combination of numerically robust NNLO Drell-Yan-plus-jet production evaluated at low transverse momentum 
cut-off~\cite{Gehrmann-DeRidder:2017mvr,Gehrmann-DeRidder:2019avi} with unresolved N$^3$LO contributions including logarithmic divergent terms~\cite{Baikov:2009bg,Lee:2010cga,Gehrmann:2010ue,Li:2016ctv,Luo:2019szz,Ebert:2020yqt,Luo:2020epw} predicted via Soft-Collinear Effective Theory (SCET)~\cite{Bauer:2000ew,Bauer:2000yr,Bauer:2001yt,Bauer:2002nz,Beneke:2002ph}. 

Our calculation for the charged-current Drell-Yan production is performed for LHC kinematics with center-of-mass energy at 13 TeV. 
The default EW parameters are based on the $G_\mu$ scheme with $M_{\mathrm Z}=91.1876$\,GeV, $\Gamma_{\mathrm Z}=2.4952$\,GeV, $G_F=1.1663787\times 10^{-5}$\,GeV$^{-2}$~\cite{ParticleDataGroup:2020ssz}. The CKM matrix is considered to be diagonal.
For each perturbative QCD order, we apply the same PDFs with $\alphas(M_{\mathrm Z})=0.118$. 
The central renormalisation ($\mu_R$) and factorisation ($\mu_F$) scales are chosen to be the invariant mass of final state Drell-Yan pair, $\mu_F=\mu_R=m_{\ell\nu}$. To estimate theoretical uncertainties, we adopt the 7-point scale variation of $\mu_F$ and $\mu_R$ by a factor of two while enforcing $1/2\le\mu_F/\mu_R\le2$.

We compare normalized differential observables computed from the central member of four sets of NNLO PDFs: NNPDF3.1~\cite{NNPDF:2017mvq}, NNPDF4.0~\cite{NNPDF:2021njg}, CT18~\cite{Hou:2019efy} and PDF2LHC21~\cite{Ball:2022hsh}. 
To describe the decay of the W boson into a charged lepton and a neutrino, we implement leading order decay matrix element with a Breit-Wigner parametrisation of the W propagator.
We use $M_{\mathrm W}=80.379$\,GeV and $\Gamma_{\mathrm W}=2.085$\,GeV reported by the Particle Data Group (PDG,~\cite{ParticleDataGroup:2020ssz}) as the default setup.
No fiducial selection criteria is applied on the final state leptons.
Two types of Breit-Wigner parametrisation are applied in this work. 
The default setup adopts the fixed-width parametrisation
\begin{equation}
\text{BW}(q^2)=-\frac{i}{q^2-M_W^2+iM_W\Gamma_W},
\end{equation}
while measurements of gauge-boson masses are usually determined with a running width parametrisation
\begin{equation}
\text{BW}(q^2)=-\frac{i}{q^2-M_W^2+iq^2\Gamma_W/M_W},
\label{eq:BWrun}
\end{equation}
where $q^2$ is the invariant mass of final state Drell-Yan pair.
We present comparisons of the default setup, running width (label `running $\Gamma_W$') and fixed width (label `fixed $\Gamma_W$') where $M_W$ and $\Gamma_W$ are transformed from on-shell to pole mass scheme according to~\cite{Bardin:1988xt}.   

\section{Results}
\begin{figure}[t]
 \centering
\includegraphics[scale=.295]{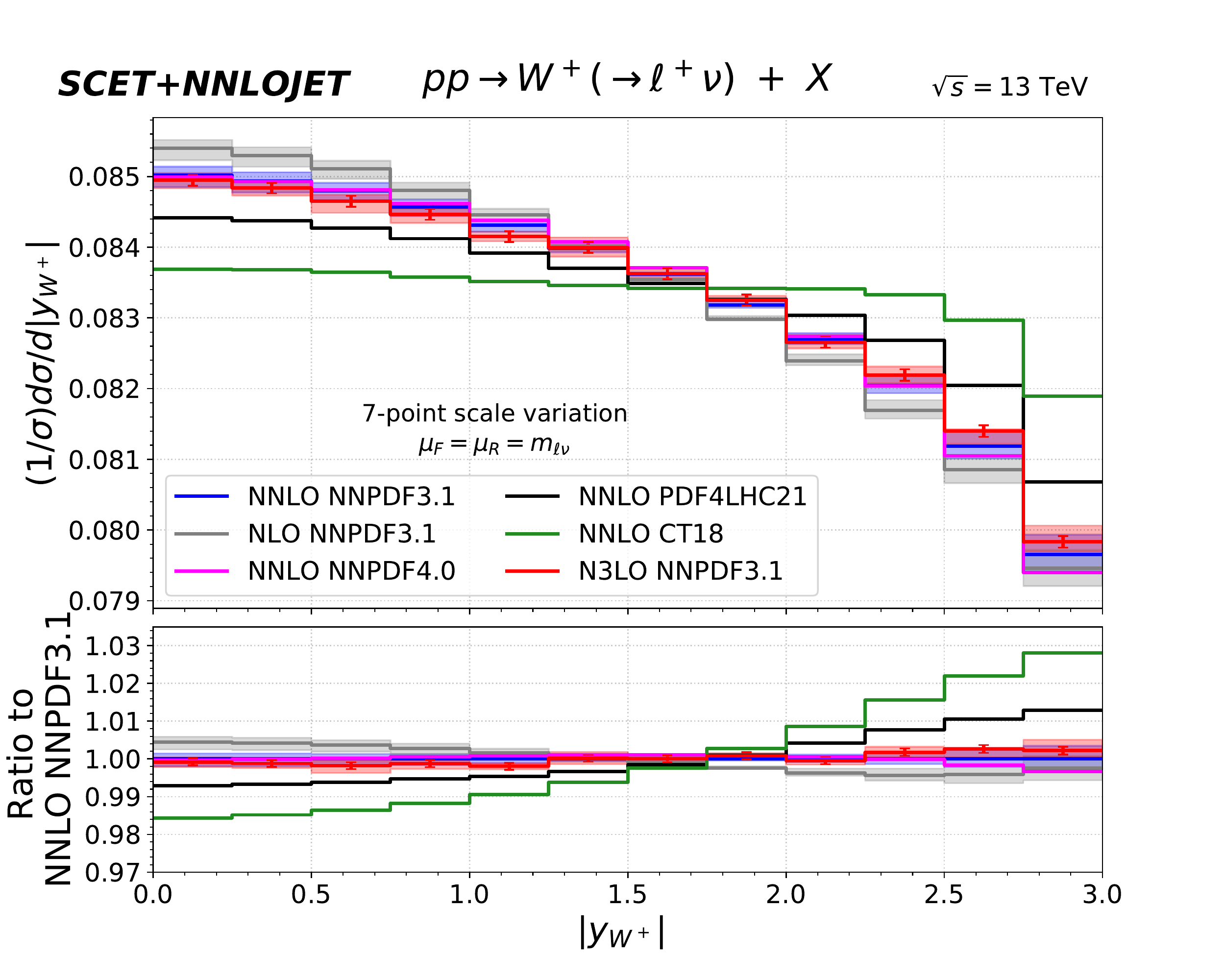}
\includegraphics[scale=.295]{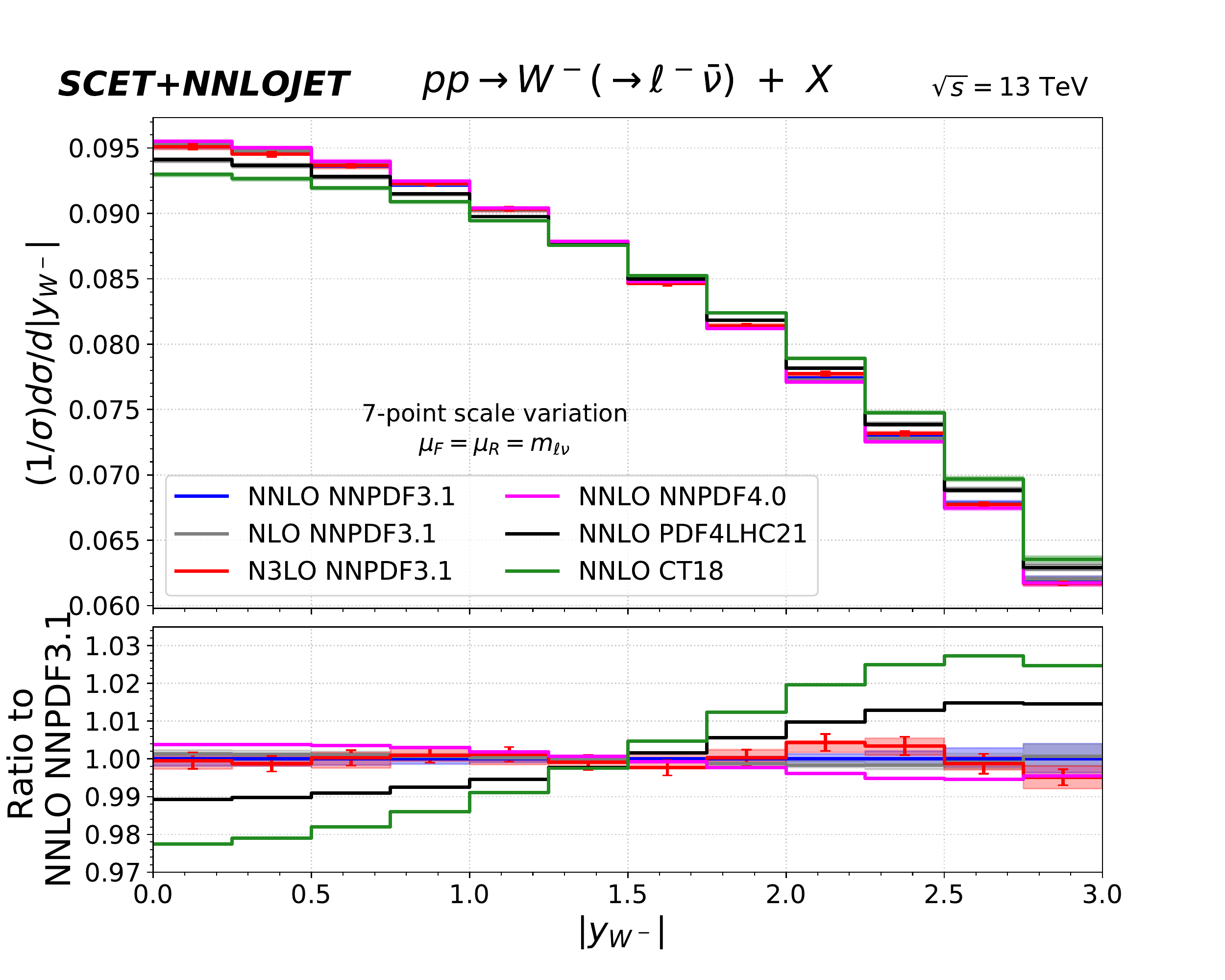}
\caption{W boson rapidity distributions with NLO to N$^3$LO QCD corrections. The NNLO corrections are calculated with four sets of PDFs. The coloured bands represent theory uncertainties from 7-point scale variation. The bottom panels show 
the ratio with respect to NNLO with NNPDF3.1 PDFs.}
\label{fig:yW}
\end{figure}

\begin{figure}[t]
 \centering
\includegraphics[scale=.295]{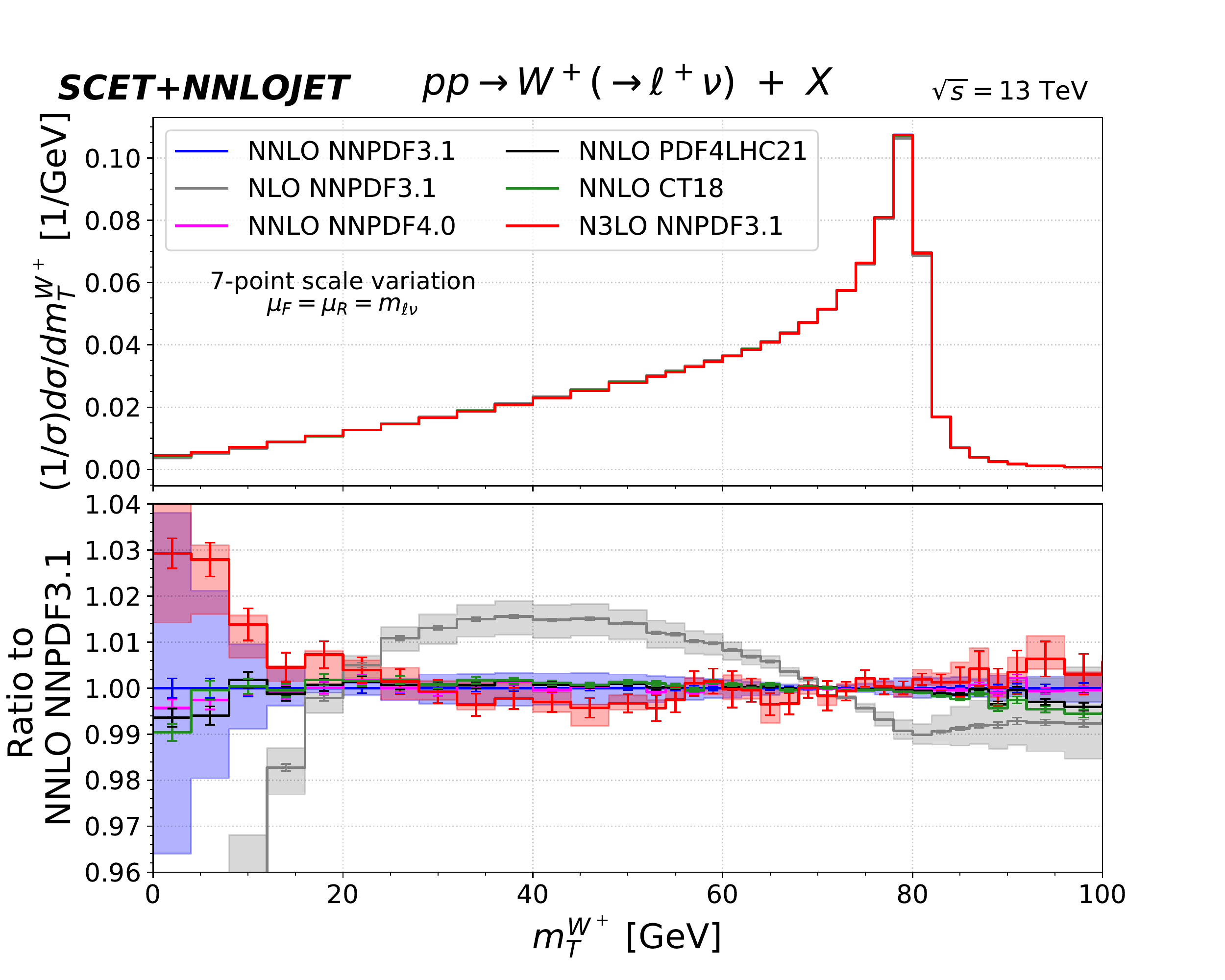}
\includegraphics[scale=.295]{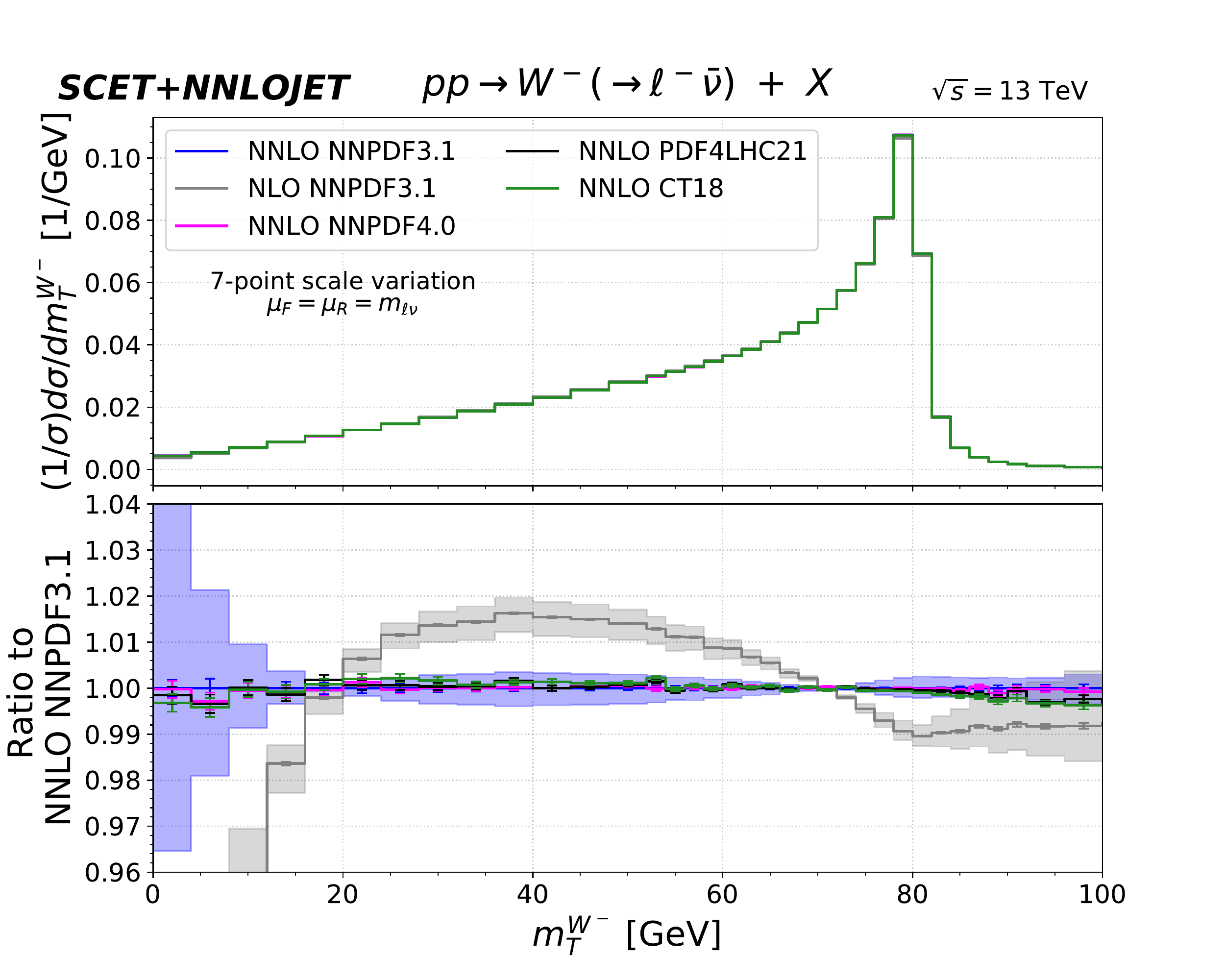}
\caption{W boson transverse mass distribution with NLO to N$^3$LO QCD corrections. The NNLO corrections are calculated with four sets of PDFs. The coloured bands represent theory uncertainties from 7-point scale variation. The bottom panel is the ratio with respect to NNLO for NNPDF3.1 PDFs.}
\label{fig:mtW}
\end{figure}

We illustrate the impact of four modern PDFs on the W boson rapidity and transverse mass distributions in Fig.~\ref{fig:yW} and ~\ref{fig:mtW}.
The rapidity distributions of the Drell-Yan pairs from ${\mathrm W}^+$ and ${\mathrm W}^-$ decays are normalized by the corresponding total cross section in Fig.~\ref{fig:yW}. 
Fixed order contributions with up to \N3\LO accuracy are illustrated with NNPDF3.1 PDFs. 
NNLO predictions with three additional PDFs (CT18, NNPDF4.0 and PDF4LHC21) are included for comparison. 
The bottom panels shows their ratio with respect to the central NNLO result from NNPDF3.1. 
The coloured bands represent theory uncertainties from the 7-point scale variation and the error bars indicate the numerical integration error at \N3\LO.
We observe a relatively large impact on the shape of distributions from the choice of PDFs, well outside scale variation bands.
For $W^+$, the central rapidity region displays about $-1.5\%$ ($-0.7\%$) deviation between CT18 (PDF4LHC21) and NNPDF3.1 PDFs, which  gradually increases 
to +2.7\% (+1.4\%) towards the forward region.
The impact of different PDF choices is stronger for $W^-$. The central rapidity region shows about $-2.3\%$ ($-1.1\%$) deviation between CT18 (PDF4LHC21) and NNPDF3.1 PDFs, which increases to +2.5\% (+1.5\%) towards the forward region.
Higher order QCD corrections at NNLO and N$^3$LO only provide small changes to the shape of rapidity distributions of less than $\pm$0.5\% throughout the plotted region. 
We notice that NNPDF3.1 and NNPDF4.0 PDFs predictions for the normalized distributions agree within $\pm$0.5\%.

In Fig.~\ref{fig:mtW} we present the normalized transverse mass distributions of charged current decay products. 
We observe similar higher order QCD corrections between $W^+$ and $W^-$ both having the dominant change to the shape of distributions from NLO to NNLO with about +$1.6\%$ around 40 GeV and $-1\%$ around the $M_W$ threshold. 
Different choices of PDFs are consistent within numerical error at NNLO accuracy through out the plotted transverse mass region.
A small negative shift about $-0.4\%$ caused by CT18 and PDF4LHC21 with respect to NNPDF3.1 (also NNPDF4.0) PDFs appears at $m_T^W$ around 100 GeV where NNLO predictions from all four PDFs are still consistent within scale uncertainty band.
The N$^3$LO QCD correction (in left panel of Fig.~\ref{fig:mtW}) agree well with NNLO predictions from various PDFs choices in the bulk of the distribution. 
It starts to have a minimum overlap in scale variation band from NNLO for $m_T^W$ below 15 GeV.
\begin{figure}[t]
 \centering
\includegraphics[scale=.325]{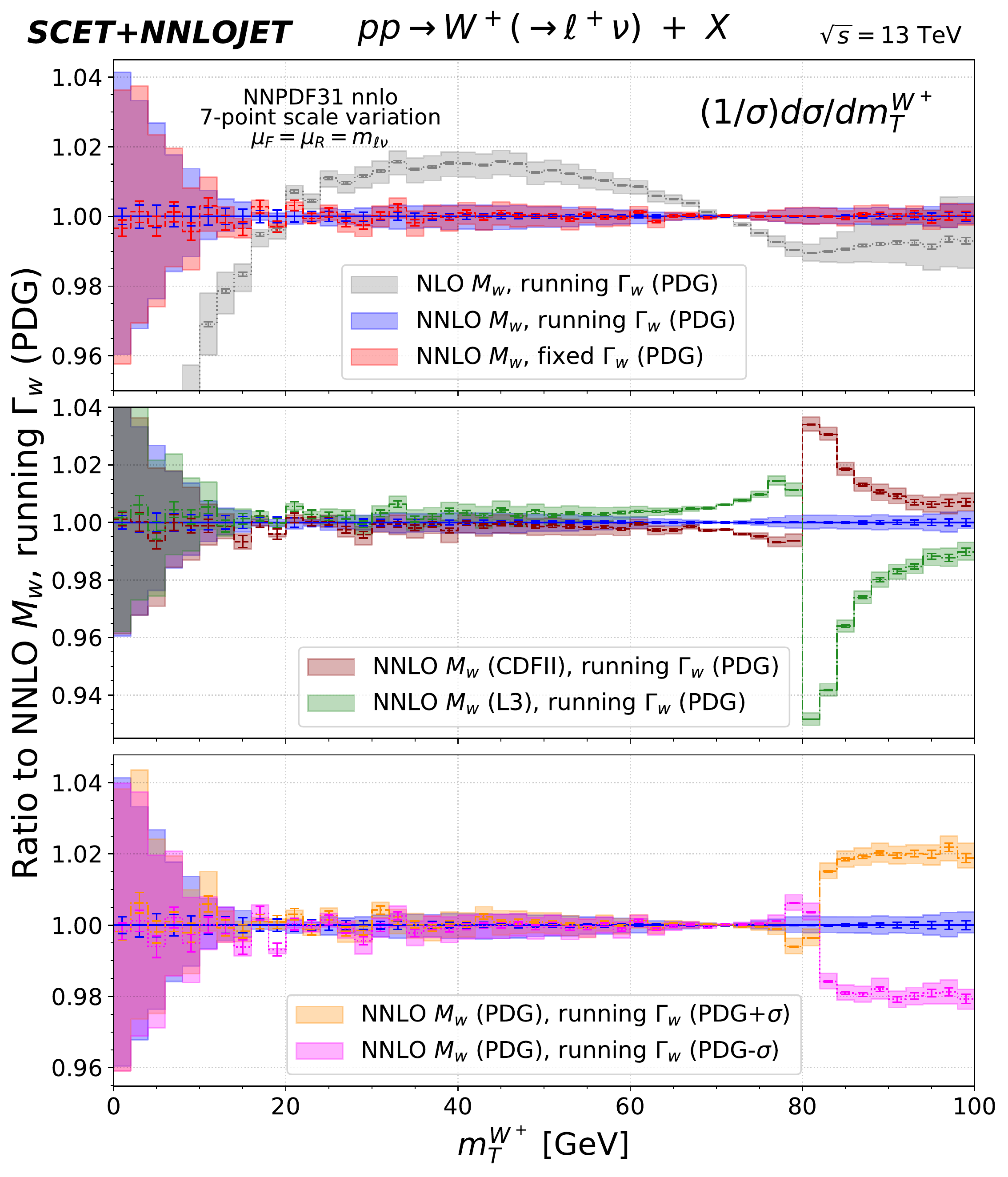}
\includegraphics[scale=.325]{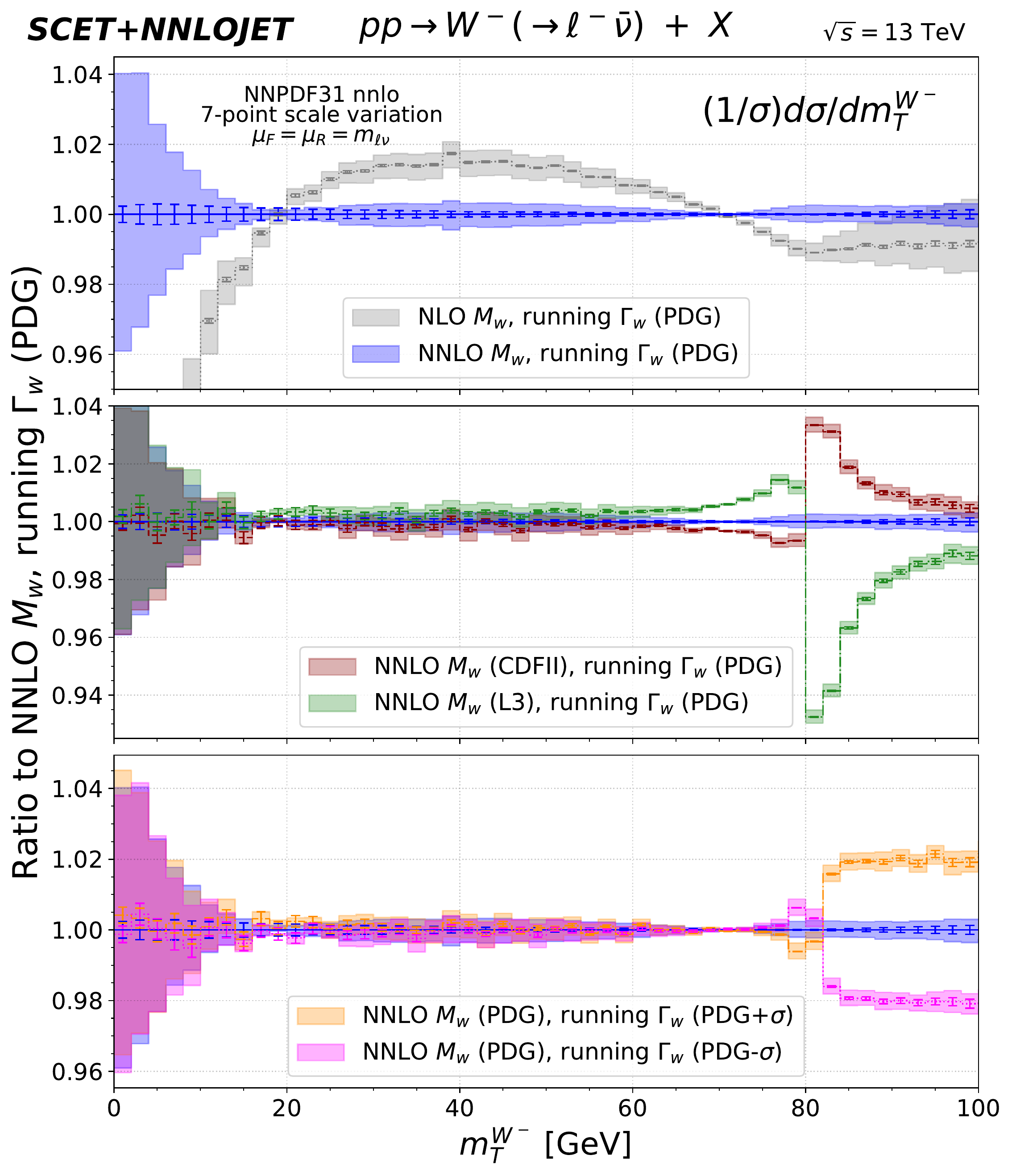}
\caption{Normalized ${\mathrm W}$ transverse mass distribution with NLO to NNLO corrections (top panel), with different $M_{\mathrm W}$ values from PDG, CDFII and L3 (middle panel) and with different $\Gamma_{\mathrm W}$ with PDG central value and $\pm\ 1\sigma$ uncertainties (bottom panel). All distributions are compared to the NNLO result with running decay width and PDG central values. Positive (negative) charged current results are in the left (right) panels. The colored bands represent theory uncertainties from 7-point scale variation.}
\label{fig:ratiomtW}
\end{figure}

We illustrate in Fig.~\ref{fig:ratiomtW} the impact of changes to the EW input parameters on the transverse mass distributions of charged current Drell-Yan production. 
Instead of the default setup, here we apply the Breit-Wigner parametrisation with running decay width described in eq.~\eqref{eq:BWrun}.
Comparing the top panels of Fig.~\ref{fig:ratiomtW} to the bottom panels of Fig.~\ref{fig:mtW}, we find excellent agreement between the two Breit-Wigner parametrisations. 
We also confirm by comparing NNLO results between running $\Gamma_W$ (in blue) and fixed $\Gamma_W$ (in red) in the top-left panel of Fig.~\ref{fig:ratiomtW} that the two parametrisations can be transformed from one to the other according to~\cite{Bardin:1988xt}.
In the middle panels of Fig.~\ref{fig:ratiomtW}, we alter the $M_W$ on-shell mass input value from the SM electroweak fit of $M_W = 80.379$ GeV to the measured results recently reported by CDFII at 80.433 GeV~\cite{CDF:2022hxs} and to the  L3 result of 80.27 GeV~\cite{L3:2005fft}. 
The weak mixing angle $\text{sin}^2\theta_W$ is a derived value in our calculation, which at tree-level is $(1 - (M_W/M_Z)^2)$.
The comparison aims to quantify the relative impact of input $M_W$ values larger or smaller than the SM electroweak fit benchmark at NNLO QCD accuracy. 
Below 60 GeV, the three chosen  $M_W$ input values yield mutually consistent results for the normalised transverse mass distribution.
We observe a strong input parameter sensitivity at the peak of the distribution around $M_W$.
Below the $M_W$ threshold, the distribution for the 
CDFII value is about 0.5\% smaller than the electroweak fit benchmark and is about 3.5\% larger right beyond the threshold.
This difference gradually reduce towards 100 GeV at about 0.7\%. 
The largest deviation between the two distributions reaches about 14 times the scale variation uncertainty.
We observe a large and negative shift beyond the threshold for the L3 input parameter, amounting to 7\% which is  about twice the difference obtained 
for  the CDFII input parameter.
In the bottom panels of Fig.~\ref{fig:ratiomtW}, we alter the $\Gamma_W$ decay width input value within the PDG uncertainty of $\pm42$ MeV. 
The impact to the normalised transverse mass distributions is minimal below 75 GeV and is stabilised at $\pm2$\% beyond the threshold compared to the benchmark distribution from the central PDG value. 
The resulting $\pm2$\% deviation corresponds to about $\pm9\ \sigma$ of the scale variation uncertainty.
Fig.~\ref{fig:ratiomtW} also demonstrates that the differences between  fixed-width and running-width schemes are minimal. 

\section{Conclusions} 
In this talk, we have presented predictions for normalised differential distributions in the charged current Drell-Yan production up to third order in perturbative QCD. 
We supplement our work in~\cite{Chen:2022lwc} to quantify the impact of PDF variations and electroweak input parameter schemes on the 
rapidity and transverse mass distributions in the charged current Drell-Yan process. 
We find nontrivial modifications to the shape of the normalised rapidity distributions at the level of $\pm 2.5\%$ in central and forward regions due to different choices of modern PDFs. Their impact is much larger than that of the N$^3$LO QCD contributions.
We compare transverse mass distributions with relative changes in the  input EW parameters with both running and fixed width parametrisation. 
Excellent agreement is observed around and beyond the $M_W$ threshold between the two parametrisations.
The shape of the transverse mass distribution is very sensitive to the input values of $M_W$ and $\Gamma_W$.
We identify regions of the transverse mass distribution where the sensitivity on  $M_W$ and $\Gamma_W$ is largest, exceeding by far the 
uncertainty from missing higher order contributions on the theory predictions.  
Our findings provide the theory input to future precision measurements of SM parameters associated with charged-current Drell-Yan production.

\section{Acknowledgments} 
We are grateful to Aude Gehrmann-De Ridder, Tom Morgan and Duncan Walker for their contributions to the $V$+jet process in the \nnlojet code. 
This work has received funding from the Swiss National Science Foundation (SNF) under contract 200020-204200 and from the European Research Council (ERC) under the European Union's Horizon 2020 research and innovation programme grant agreement 101019620 (ERC Advanced Grant TOPUP). 
This work is also supported in part by the UK Science and Technology Facilities Council (STFC) through grant ST/T001011/1 and in part by the Deutsche Forschungsgemeinschaft (DFG, German Research Foundation) under grant 396021762-TRR 257. 
H. X. Z. is supported by the Natural Science Foundation of China (NSFC) under contract No. 11975200.

\end{document}